\documentclass[12pt]{article}
\usepackage{amssymb,amsmath,cite,bm}
\usepackage[dvips]{graphicx,color}
\usepackage{psfrag,subfigure}
\begin{document}
\title{\bf Frictional Impacts in Multibody Systems}

\author{Farhad Aghili\thanks{email: faghili@encs.concordia.ca}}

\date{}
\maketitle

\begin{abstract}
A unifying slipping and sticking frictional impact model for multibody systems in contact with a frictional surface is presented. It is shown that the model can lead to energetic consistency  in both slip state  and stick state upon imposing specific constraints on  the coefficient of friction (CoF) and the coefficient of restitution (CoR). A discriminator in the form of a quadratic function of the pre-impact velocity is introduced based on isotropic Coulomb constraint such that its sign determines whether the impact occurs in the sticking mode or in the slipping mode just prior to the contact. Solving the zero-crossings of such a function in terms of the CoF and the CoR variables leads to another discriminator called Critical CoF, which is the lowest static CoF required to prevent the subsequent impulse vector violating the isotropic friction cone constraint.  Investigating conditions for the energetically consistent impact model  reveals that the maximum values of either CoR or CoF should be limited depending on stick state or slip state. Furthermore, it is shown that these upper-bound limits in conjunction with the introduced Critical CoF variable can be used to specify the admissible set of CoR and CoF parameters, which can be represented by two distinct regions in the plan of CoF versus CoR. 
\end{abstract}

\section{Introduction}

Nonlinear dynamics arises from slipping and sticking frictional impact phenomena occurs in many robotics applications involving multiple contacts or formations  of closed-loop topology \cite{Muller-1995,Aghili-2005,McClamorch-Wang-1988,GarciadeJalon-Bayo-1994,Blajer-Schiehlen-Schirm-1994,Aghili-2020a,Preclik-2014,Aghili-Su-2017}. Examples include industrial manipulators performing complex contact tasks \cite{Gottschlich-Kak-1989,Walker-1994,Aghili-2010h}, force control of constrained robots  \cite{Dupree-Liang-2008,Su-Stepanenko-1995,Aghili-Buehler-Hollerbach-1997a,Zhang-Xi-2009,Aghili-2015b}, walking robots \cite{Marhefka-Orin-1999,Mu-Wu-2006,Konno-Myojin-2011,Yoshida-Takeuchi-2014} and space robotics capturing free-floating objects \cite{Nenchev-Yoshida-1998}. Such systems can be generally treated as multibody systems (MBSs) with a time-variant structure, and hence  they often exhibit a nonsmooth behaviour due to friction, unilateral constraint, and impact~\cite{Keller-1986,Glocker-Pfeiffer-1995,Stronge-James-Ravani-2001,Johansson-2001,Pfeiffer-2009,Aghili-2020a}. Consequently, there are two main challenges in dynamics formulation of these systems: (i) Discontinuity due  to activation and deactivation of unilateral constraints and dynamics itself, (ii) friction in the contact, which causes fundamental change in the dynamic behavior of the systems when the friction characteristic changes from stick state to the slip state and vice versa.

Frictional impact is often modeled as instantaneous events because  the system's velocities are of bounded variation and hence they cannot be assumed continuous  during impact events. Therefore, the system's post-impact states cannot be calculated from an acceleration model. On the contrary, they have to be derived from an impact model by incorporating the impulse-momentum balance of the entire system together with a restitution law and a friction law. Therefore, modeling frictional impacts demand proper combination of a restitution law and friction law.  The Newtonian restitution law is extensively used  to model the normal direction supplemented by the Coulomb friction law in the tangential direction frictional impacts \cite{Glocker-2013}. According to isotropic Coulomb friction,  frictional impacts  occur in the sticking mode if the magnitudes of the tangential and normal impulses do not violate the friction cone constraint, otherwise the frictional impacts occur in the slipping mode. Impact dynamics fundamentally changes when the MBS switches from sticking mode to slipping mode or vice versa. That is why the properties of the MBS during impact, such as its energy change during impact, depend on whether the system is in slip state or stick state. Simulation and analysis of frictional impact in MBSs calls for a unifying formulation which is energetically consistent, i.e. the model is always  dissipative or energy preserving, for both slip state and stick state. Ideally, an unified formulation has to be general enough to give a closed-form and energetically-consistent solution for the velocity jump and the subsequent impulse in slipping and sticking frictional impacts.

Contact dynamics was formulated by Moreau \cite{Moreau-1988} through a non-smooth mechanics  modelling approach, which  allows the time evolutions of the positions and of the velocities to be non-smooth and discontinuous. A good survey of impact analysis in the framework of nonsmooth mechanics can be found in \cite{Brogliato-1999}. Numerical methods for nonsmooth dynamical systems applicable to both
mechanics and electronics are described in \cite{Acary-Brogliato-2008}. Stability theory is developed in \cite{Leine-2008} for non-smooth dynamical models, which arise from  mechanical systems with unilateral constraints, such as unilateral contact, impact and friction. The frictional contact problem is cast in a mixed penalty-duality formulation, known as the prox formulation, by Alart et al. \cite{Alart-Curnier-1991} as an alternative to the well-known linear and nonlinear complementarity problem  formulations. However, it was later shown  in \cite{Schindler-Nguyen-2011} that these two formulations are indeed equivalent. A unified theory and application of dynamical systems that incorporate  hard inequality constraint such as mechanical systems with impact or electrical circuits with diodes are comprehensively addressed in \cite{Stewart-2011}. A new friction model based on a regularization of the Coulomb friction model  is incorporated in the collision micro-dynamics hard contact models for simulation of the macroscopic behaviour of granular matter \cite{Preclik-2014}.  The earliest analytical investigation into  robot dynamics during impact based on modelling of collisions between the robot and the environment was reported in \cite{Zheng-Hemami-1985}. It followed by design and evaluation of impact control schemes for robots during collisions \cite{Volpe-Khosla-1993,Walker-1994}. In the literature, the primary approach to modelling frictional impact in MBSs is based on combining the equation of momentum-balance together with a restitution and friction laws \cite{Glocker-Pfeiffer-1995,Stronge-2000}. In this approach, the impact is assumed to be an infinitesimal event and subsequently the set of the impulse-momentum equations are algebraically solved to produce velocity jump rather than updating the velocity from  integration of the acceleration vector. In this formulation, the impact is characterized by the coefficient of restitution (CoR), which is defined as the ratio of the local velocities after and before collision. Other approaches assume smooth compliant modeling of impact where  the impulsive force is typically presented by a linear or nonlinear spring-damper model \cite{Khulief-2013,Hunt-Crossley-1975}, e.g., the Haunt-Crossley nonlinear spring-dashpot model. However, it has been established that   a linear or non-linear  compliant model becomes equivalent to the momentum-balance model if the coefficient of restitution is specifically selected according to the damping and stiffness properties of the compliance model \cite{Goldsmith-1960,Lankarani-Nikravesh-1990,Marhefka-Orin-1999}. It is known that  frictionless impacts for single or multiple contacts based on Newton's restitution law are always dissipative \cite{Aghili-2019a,Glocker-2013,Kane-Levinson-1985} if a global CoR is selected to be  within $[0, \; 1]$. However, the  Newton's restitution law may potentially produce energetically inconsistent results  when combined with friction in some contact situations  \cite{Kane-Levinson-1985,Wang-Mason-1992,Djerassi-2009,Glocker-2013}. Glocker et al. \cite{Glocker-Pfeiffer-1995} introduced an impact model for two-dimensional
contacts under Coulomb friction and Poisson’s hypothesis
that is based on a linear complementarity formulation. Although this formulation of impact has been designed to be energy dissipative, it does not correspond to fundamental physical properties \cite{Jia-2013,Chatterjee-Ruina-1988}, the normal impulse may sometimes be too small to prevent penetration yielding an unrealistic
solution \cite{Glocker-Pfeiffer-1995}, and finally, likewise Newton's impact law, Poisson's hypothesis may predict
an increase in the kinetic energy  \cite{Wang-Mason-1992,Stewart-Trinkle-1996,Jia-2013}.
Comparisons between  different impact laws in terms of their respective abilities to correctly model dissipation and dispersion of energy are shown in \cite{Nguyen-Brogliato-2013}.
The Kane's example in which a double pendulum strikes a flat surface by a frictional impact has been thoroughly analyzed  in \cite{Glocker-2013} by decomposing the friction elements into its basic primitives. This technique allowed to reveal the mechanism leading to energy inconsistency in the Kane's example. The Poisson's and Coulomb law are combined in \cite{Wang-Mason-1992} to solve the impact problem at force level for  two-dimensional single contact cases. A dissipation principle for resolving post-impact tangential velocities after simultaneous
oblique impact events on a MBS is proposed in \cite{Flickinger-Bowling-2009}. The proposed dissipation principle allows for changes in the dynamic coefficient of friction depending on the orientation of the impacting surfaces.

In this paper,  a unified frictional impact model which is consistent in both slip state and stick state along with comprehensive analysis of energetic consistency leading to admissible values of CoR and CoF are presented. Derivation of the unified model is made possible by introducing a quadratic function of pre-impact velocity, the sign of which determines whether a frictional impact occurs in slip state or in stick state. It will be shown that zero-crossings of such a implicit function can be  solved in terms of a new variable, called Critical CoF, which is an explicit function of CoR. It turns out that the Critical CoF is a convenient discriminator function, which determines the minimum required static CoF to prevent slipping during an impact. The condition for energetically consistent description of the impact model for both stick state and slip state is  comprehensively investigated leading to identification of two exclusive regions in the plan of CoR and CoF variables, where a MBS becomes energetically consistent during slip and stick frictional impact. Finally, a case study along with simulation results underpins the impact model and the analytical results.

\section{Dynamics model}

Consider a frictional contact surface with static and dynamic coefficients of friction  $\mu_s$ and $\mu_d$. The vector of contact force
\[ \bm\lambda = \begin{bmatrix} \lambda_n \\ \bm\lambda_t \end{bmatrix} \]
consists of normal force $\lambda_n \in \mathbb{R}$ and  tangential force $\bm\lambda_t \in \mathbb{R}^2$. The contact friction has two modes: sticking and slipping. The sticking refers to a situation where the magnitude of the tangential force in the contact is not sufficient to overcome the static friction and hence causing the relative motion in the contact, i.e.,
\begin{equation} \label{eq:friction_cone}
\| \bm\lambda_t \| \leq \mu_s \lambda_n.
\end{equation}
Otherwise, slipping friction force  occurs where the friction force vector has the magnitude of normal force times $\mu_d$ and it opposes the vector of tangential velocity $\bm v_t$. Inequality  \eqref{eq:friction_cone} is called friction cone constraint and can be equivalently transcribed by the following quadratic inequality on the entire vector of contact force: $\bm\lambda^T \bm U \bm\lambda \geq 0$, where
\begin{equation} \label{eq:Udef}
\bm U=\mbox{diag}(\mu_s^2, \; -1, \; -1).
\end{equation}
From the preceding discussion, we can generalize the description of the contact force vector in a frictional surface by
\begin{equation}
\bm\lambda = \left\{ \begin{array}{ll} \begin{bmatrix} \lambda_n \\ \bm\lambda_t \end{bmatrix} \qquad & \mbox{if} \quad   \bm\lambda^T \bm U \bm\lambda \geq 0  \\
 \begin{bmatrix}  \lambda_n \\ - \mu_d \frac{\bm v_t}{\| \bm v_t \|} \lambda_n  \end{bmatrix}  & \mbox{otherwise} \end{array} \right.
\end{equation}
Now, consider a multibody system  with generalized coordinate $\bm q \in \mathbb{R}^n$  subject to the surface constraint with frictional contact. Dynamics equation of such system in the {\em sticking friction mode}  can be described by
\begin{subequations} \label{eq:accleration_sticking}
\begin{align}
&\bm M(\bm q) \ddot{\bm q} + \bm h  = \bm A^T \bm\lambda \\
\mbox{subject to:} \quad & \bm A \dot{\bm q} = \bm 0 \\ \label{eq:LUL}
&  \bm\lambda^T \bm U \bm\lambda \geq 0 \\
&  \lambda_n \geq 0.
\end{align}
\end{subequations}
Where ${\bm  M}(\bm q) \in \mathbb{R}^{n \times n}$  is the inertia matrix;
vector ${\bm h} \in\mathbb{R}^{n}$ contains all nonlinear terms plus forces due to gravity and actuators, and $\bm A=[\bm a_n^T \; \bm A_t^T ]^T \in \mathbb{R}^{3 \times n}$ is the overall Jacobian matrix, i.e., $\bm a_n\in \mathbb{R}^{1  \times n}$ and $\bm A_t\in \mathbb{R}^{2 \times n}$ are associated with the normal force and tangential force, respectively. Hereafter, the normal contact force is treated as unilateral force  whereas the  tangential contact force is  treated as bilateral force if the contact force remains to be inside or on the friction cone constraint. That is, we assume no slipping occurs in the contact and subsequently the tangential contact is treated as bilateral constraint. In other words, slipping in the contact occurs if the friction cone constraint \eqref{eq:LUL} does not hold. Then  according to the Coulomb's friction law, the magnitude of slipping friction is equal to $\mu_d \lambda_n$ and it opposes the direction of tangential velocity $\bm v_t=\bm A_t \dot{\bm q}$. Therefore, dynamics equation of the system in the {\em slipping friction mode}  can be described by
\begin{subequations} \label{eq:accleration_slipping}
\begin{align} \label{Mddq_slipping}
&\bm M(\bm q) \ddot{\bm q} + \bm h  = \big( \bm a_n^T - \mu_d \bm b^T \big) \lambda_n \\
\mbox{subject to:} \quad & \bm a_n \dot{\bm q} = \bm 0 \\
&  \lambda_n \geq 0
\end{align}
\end{subequations}
where
\begin{equation}
\bm b(\bm q, \dot{\bm q}) =  \frac{1}{\| {\bm A_t} \dot{\bm q} \|} \dot{\bm q}^T \bm A_t^T \bm A_t.
\end{equation}
It is worth nothing that $\bm a_n - \mu_d \bm b$ in \eqref{Mddq_slipping} can be treated as artificial Jacobian of the constrained system in the slip state.

Transition from noncontact phase to contact phase causes impact accompanied by impulsive forces and sudden change in the generalized velocity. Due to discontinuity of the velocities during an impact event, the velocity  instantaneously jumps and thus requiring infinitely large acceleration and constraint force. Therefore, the acceleration models  \eqref{eq:accleration_sticking} or \eqref{eq:accleration_slipping} are not adequate to determine post-impact velocities rather an impact model is required to deal with the impulsive constraint force and discontinuities in the velocities.  Since the impact is assumed to be an infinitesimal event, it is also reasonable to assume either stick friction or slip friction occurs during the impact. That is either  \eqref{eq:accleration_sticking} or \eqref{eq:accleration_slipping} governs the impact dynamics. Although it is not possible to calculate the velocity change at the time of impact through integration of the equations of motion, it is possible to calculate the velocity change using the {\em Newton's impact law}. Suppose impacts  occur at time interval $[t^-, \; t^+]$, where the impact duration $\delta t= t^+ - t^-$ is infinitesimal. Since the generalized coordinate $\bm q$ is assumed to be constant over the impact, the mass matrix $\bm M(\bm q)$ and the Jacobian $\bm A (\bm q)$ remain unchanged during  the impact. Therefore, one can carry out integration of the differential equation \eqref{eq:accleration_sticking} over $[t^-, \; t^+]$ to obtain impact equation of the system as
\begin{equation} \label{eq:sticking_dirac}
\bm M(\dot{\bm q}^+ - \dot{\bm q}^-) = \bm A^T \bm i.
\end{equation}
Here, $\bm i$ is the impact or the Dirac integral of the contact forces, i.e.,
\begin{equation}
\bm i  = \lim_{\delta t \rightarrow 0} \int_{t^-}^{t^- + \delta t} \bm\lambda \; {\rm d}t,
\end{equation}
and $\dot{\bm q}^- = \dot{\bm q}(t^-)$ and $\dot{\bm q}^+ = \dot{\bm q}(t^+)$ are pre-impact and post-impact velocities.  Notice that in derivation of \eqref{eq:sticking_dirac}, we assumed  $\bm h$ to be a continuous function consisting of nonimpulsive terms, and thus it vanishes by the integration. The velocity jump can be captured by the  restitution model
\begin{equation} \label{eq:Edq}
\bm A \dot{\bm q}^+ = - \bm E \bm A \dot{\bm q}^-,
\end{equation}
where the restitution matrix. An impact law of Newton's type imposes a kinematic condition on the impact via scalar restitution $e$ on the pre- and post-impact relative velocities \cite{Glocker-2013}. The restitution matrix can be written in a general form   $\bm E =\mbox{diag}[e_n, e_t, e_t]$, where $e_n$ and $e_t$ are CoRs associated with normal and tangential impacts. Tangential compliance is typically negligible at contacts between
smooth and hard materials such as steel and glass, while it may become
more prominent with materials such as rubber, clay, and wood \cite{Jia-2013}. Therefore, the restitution matrix typically takes the form $\bm E =\mbox{diag}[e, 0, 0]$ with $e$ being the coefficient of restitution \cite{Glocker-2013}. It is worth mentioning that the CoR can be derived from a linear or nonlinear spring-dashpot model or Hertz's model to make CoR physically sound \cite{Goldsmith-1960, Hunt-Crossley-1975,Marhefka-Orin-1999,Lankarani-Nikravesh-1990,Nguyen-Brogliato-2013}. For instance, it has been shown that the at low impact velocity and for materials having linear elastic behaviour the CoR can be effectively approximated by the equation $e = 1 - \alpha v_n$, where parameter $\alpha=2c/3k$ with $c$ and $k$ being the damping and stiffness constants of the spring-dashpot model and $v_n$ being the relative normal velocity \cite{Hunt-Crossley-1975,Marhefka-Orin-1999,Lankarani-Nikravesh-1990}.
The equations of momentum balance \eqref{eq:sticking_dirac} combined with the restitution equation \eqref{eq:Edq} are sufficient to solve for the vectors of post-impact velocity and impulse. To this end, by pre-multiplying  both sides of \eqref{eq:sticking_dirac} by $\bm A \bm M^{-1}$ we get:
\begin{equation} \label{eq:AMATi}
\bm A \dot{\bm q}^+ - \bm A\dot{\bm q}^- = \bm A \bm M^{-1}\bm A^T \bm i
\end{equation}
Substituting $\bm A \dot{\bm q}^+ $ from \eqref{eq:Edq} into \eqref{eq:AMATi} and then solving the resultant equation for  $\bm i$, we arrive at
\begin{subequations} \label{eq_sticking_mode}
\begin{equation} \label{eq:impact_sticking}
\bm i = - \bm G (\bm E + \bm I) \bm A \dot{\bm q}^-,
\end{equation}
where $\bm G= (\bm A \bm M^{-1} \bm A^T)^{-1}$. Finally, upon substitution of the expression of impulse from \eqref{eq:impact_sticking} into \eqref{eq:AMATi}, we can write the expression of the post-impact velocity by
\begin{equation} \label{postvel_striction}
\dot{\bm q}^+ = \dot{\bm q}^- - \bm M^{-1} \bm A^T \bm G (\bm I + \bm E)\bm A \dot{\bm q}^-.
\end{equation}
\end{subequations}
An isotropic Coulomb constraint restricts the magnitude of the friction impulse by the following inequality
\begin{equation} \label{eq:i_quadratic}
\bm i^T \bm U  \bm i \geq 0.
\end{equation}
Upon substitution of $\bm i$ from \eqref{eq:impact_sticking} into \eqref{eq:i_quadratic}, the latter inequality can be equivalently written in terms of the following {\em discriminator function} of pre-impact velocity
\begin{equation} \label{eq:sigma}
\sigma:=\dot{\bm q}^{-T} \bm Q \dot{\bm q}^- \geq 0,
\end{equation}
where the matrix is defined as
\begin{equation} \label{eq:Qdef}
\bm Q:=\bm A^T \bm G (\bm I + \bm E) \bm U (\bm I + \bm E)\bm G \bm A.
\end{equation}
In other words, if the pre-impact velocity satisfies $\sigma \geq 0$, then the impact is with sticking fiction mode and hence  solutions \eqref{eq_sticking_mode} is valid. Otherwise, the frictional impact involves slipping and then the solution should take the following steps. Suppose $\bm i=[i_n \; \; \bm i_t^T]^T$, where $i_n \in \mathbb{R}$ and $\bm i_t \in \mathbb{R}^2$ are normal and tangential impulses. Then, integration of equation \eqref{Mddq_slipping} gives the governing slipping impact model as
\begin{equation} \label{eq:momentum_slipping}
\bm M(\dot{\bm q}^+ - \dot{\bm q}^-) = (\bm a_n^T - \mu_d \bar{\bm b}^T ) i_{_n},
\end{equation}
where
\begin{equation} \label{eq:b_bar i}
\int_{t^-}^{t^- + \delta t} \bm b(\bm q , \dot{\bm q}) \lambda_n \; {\rm d} t =  \bar{\bm b} i_{n}
\end{equation}
and $\bar{\bm b} = \bm b(\bm q , \bar{\dot{\bm q}})$ with $\bar{\dot{\bm q}}$ being the average velocity during the slipping impact. Notice that the computation in \eqref{eq:b_bar i} does not assume the slip velocity being constant during the impact, rather $\bar{\bm b}$ is computed based on average of pre- and post-impact velocities. In other words, we assume the average value of $\bm b$ calculated at the average velocity $\bar{\dot{\bm q}}$ in order to be able to evaluate the integral in \eqref{eq:b_bar i}. Alternatively, one may use one of the proposed trajectories  during the compression and expansion phases of an impact \cite{Pfeiffer-2009} to find the integration of $\bm b$ over the short time of impact. The restitution model in the case of slipping friction mode is simply described by the following equation
\begin{equation} \label{eq:restitution_slipping}
\bm a_n \dot{\bm q}^+ = - e \bm a_n \dot{\bm q}^-
\end{equation}
In a development similar to \eqref{eq:AMATi}-\eqref{postvel_striction}, one can solve  equations \eqref{eq:momentum_slipping} and \eqref{eq:restitution_slipping} for
\begin{subequations} \label{eq_slipping_mode}
\begin{align} \label{eq:in_slipping}
i_{n} &= -(1+e) g \bm a_n \dot{\bm q}^-, \\ \label{eq:velocity_postimpact}
\dot{\bm q}^+ &= \dot{\bm q}^- - (1+e) g \bm M^{-1}(\bm a_n^T \bm a_n - \mu_d \bar{\bm b}^T \bm a_n) \dot{\bm q}^-,
\end{align}
where $g=1/\big( \bm a_n \bm M^{-1} \bm a_n^T - \mu_d \bm a_n \bm M^{-1} \bar{\bm b}^T \big)$. On the other hand, the tangential impact in the case of slip state is related to normal impact by
\begin{equation}\label{eq:it_slipping}
\bm i_{t} = - \mu_d \frac{\bm A_t \bar{\dot{\bm q}}}{\| \bm A_t \bar{\dot{\bm q}} \|} i_{n}.
\end{equation}
Thus, from \eqref{eq:in_slipping} and \eqref{eq:it_slipping}, we can describe the overall vector of impulse by
\begin{equation} \label{eq:impulse_slipping}
\bm i = -(1+e) g \begin{bmatrix} \bm a_n \\  - \mu_d \frac{\bm A_t \bar{\dot{\bm q}}}{\| \bm A_t \bar{\dot{\bm q}} \|} \bm a_n \end{bmatrix} \dot{\bm q}^-
\end{equation}
\end{subequations}
It is important be pointed out that solution \eqref{eq:in_slipping} is kinematically consistent if
\begin{equation}
i_n \geq 0
\end{equation}
Since  the pre-impact normal velocity is always negative, i.e.,  $ v_n^- =\bm a_n \dot{\bm q}^- <0$, one can conclude from expression \eqref{eq:in_slipping} that $i_n \geq 0$ is held if and only if $g>0$. In other words, the kinematically consistency is tantamount to satisfying the following inequality
\begin{equation} \label{eq:g>0}
{\bm a_n \bm M^{-1} \bm a_n^T} - \mu_d {\bm a_n \bm M^{-1} \bm A_t^T \hat{\bm u} } > 0
\end{equation}
where $\hat{\bm u}=\frac{\bm A_t \bar{\dot{\bm q}}}{\| \bm A_t \bar{\dot{\bm q}} \|}$ is the unit directional vector along with the slipping velocity. It will be shown later in Section~\ref{sec:energetic_consistency} that the kinematically consistency condition is tantamount to energetic consistency of impacts.

In summary, computation of sticking/slipping frictional impact may proceed as follows:
\begin{enumerate}
\item For a given pre-impact velocity, compute the discriminator $\sigma=\dot{\bm q}^{-T} \bm Q \dot{\bm q}^-$ to check the subsequent inequality condition;
\item if $\sigma \geq 0$, then use the set of equations \eqref{eq_sticking_mode} to determine the post-impact velocity and impulsive force;
\item if $\sigma < 0$, then use set of equations \eqref{eq_slipping_mode} to determine the post-impact velocity and impulsive force.
\end{enumerate}

\subsection{Critical coefficient of friction}
It is evident from \eqref{eq:Udef}, \eqref{eq:sigma}, and \eqref{eq:Qdef} that the discriminator function $\sigma$ is a quadratic function of static CoF, CoR, and pre-impact velocity, i.e., $\sigma=\sigma(\mu_s, e, \dot{\bm q}^-)$. A natural question arises that what values of the CoF will make $\sigma=0$. Let us define the Critical CoF, denoted by $\mu_{\rm cr}$, as the root of the discriminator function, i.e.,
\begin{equation} \label{eq:mu_cr}
\sigma(\mu_{\rm cr},e, \dot{\bm q}^-)=0.
\end{equation}
Then, the variable $\mu_{\rm cr}$ must be a function of CoR and pre-impact velocity, i.e., $\mu_{\rm cr}= \mu_{\rm cr}(\dot{\bm q}^-,e)$. Defining auxiliary vector $\bm\xi=[\xi_n \;\; \bm\xi_t^T]^T=\bm G \bm A \dot{\bm q}^-$, we can write the quadratic expression of $\sigma$ in \eqref{eq:sigma} in terms of the elements of vector $\bm\xi$ as follows
\begin{equation} \label{eq:sigma_mu2}
\sigma = \mu_s^2 \xi_n^2(e+1)^2 - \| \bm\xi_t \|^2,
\end{equation}
Then, the valid solution to t quadratic equation $\sigma=0$ takes the form
\begin{equation} \label{eq:mucr_e}
\mu_{\rm cr}=\frac{a}{ e+1}, \qquad \mbox{where} \quad a=\| \bm\xi_t \|/|\xi_n|.
\end{equation}
It is clear from the above second-order polynomial that $\mu_s \geq \mu_{\rm cr}$ implies $\sigma \geq 0$ and conversely $\mu_s < \mu_{\rm cr}$ implies $\sigma < 0$. In other words, $\mu_{\rm cr}$ is also a discriminator which determines  the lowest static CoF required to prevent slipping during  impacts. That is to say
\begin{equation} \label{eq:mucr_sticking_slipping}
\left\{ \begin{array}{ll} \mu_s \geq \mu_{\rm cr} \qquad & \mbox{sticking  impact} \\
\mu_s < \mu_{\rm cr}  & \mbox{slipping impact}  \end{array} \right.
\end{equation}
It worths noting that there is always an algebraic solution for the critical CoF. However, from a physics point  of view, the  possibility of CoF being greater than 1 is not very practical. In that case, one can conclude from  \eqref{eq:mucr_e} that if
\begin{equation}
a>1 \quad \wedge \quad e< a-1
\end{equation}
then there is no physically feasible solution for the critical CoF, meaning that slipping impact can not be avoided for all CoF inside $[0, \; 1)$.

\subsection{Energetic consistency} \label{sec:energetic_consistency}
Energy lost during an impact is an important quantity not only to gain insight into complex physical phenomenon during contact but to examine whether an impact model is physically consistent \cite{Stronge-1991,Stoianovici-Hurmuzlu-1996}.
The energy dissipation done by the contact force during impacts can be expressed as
\begin{equation} \label{eq:W_loss}
W_{\rm loss} =  \int_{t^-}^{t^- + \delta t} \bm v \cdot \bm\lambda \; {\rm d} t =    \bar{\bm v}  \cdot \bm i
\end{equation}
where $\bar{\bm v} =\frac{1}{2}(\bm v^+ + \bm v^-)$ is the average velocity during the contact.  In the following analysis, we compute the equation of power \eqref{eq:W_loss} and drive the conditions for its negativity for two district modes of frictional impact, i.e., the sticking mode where  $\mu \geq \mu_{\rm cr} $ and slipping mode where $\mu < \mu_{\rm cr}$.

\subsubsection{Sticking mode}

Consider stick frictional impacts: From the restitution law $\bm v^+ = - \bm E \bm v^- $  we can write the expression of average contact velocity as
\begin{equation} \label{eq:v_average}
\bar{\bm v}=\frac{1}{2}(\bm I - \bm E) \bm A \dot{\bm q}^- \quad  \Leftarrow \quad  \mu_s \geq \mu_{\rm cr}
\end{equation}
Substituting expression of $\bar{\bm v}$ and $\bm i$ from  \eqref{eq:v_average}  and \eqref{eq:impact_sticking} into \eqref{eq:W_loss} yields
\begin{align} \notag
W_{\rm loss} &=   -\frac{1}{2}  \dot{\bm q}^{-T} \bm A^T (\bm I - \bm E) \bm G (\bm I + \bm E) \bm A  \dot{\bm q}^- \\ \label{eq:loss_sticking}
&=  -\frac{1}{2}  \dot{\bm q}^{-T} \bm A^T (\bm G - \bm E \bm G \bm E) \bm A  \dot{\bm q}^- \quad  \Leftarrow \quad  \mu_s \geq \mu_{\rm cr}
\end{align}
in which \eqref{eq:loss_sticking} is obtained by using the fact that $\bm G \bm E - (\bm G \bm E)^T$ is a skew symmetric matrix and hence the corresponding quadratic term is eliminated in expression of RHS of \eqref{eq:loss_sticking}. Therefore, one can infer from \eqref{eq:loss_sticking} that a sticking frictional impact is energetically consistent if the value of the CoR is upper-bounded as
\begin{equation} \label{eq:e<e_star}
e \leq e^*,
\end{equation}
where $e^*$ is root of the following quadratic  function
\begin{equation} \label{label:quadratic}
{\bm v}^{-T} \big(\bm G - \bm E(e^*) \bm G \bm E(e^*) \big) {\bm v}^- = 0.
\end{equation}
Alteratively,  one can conclude from \eqref{eq:loss_sticking} that the power loss during sticking frictional impact is always negative semidefinite if the following symmetric matrix is positive semidefinite
\begin{equation}
\bm G - \bm E \bm G \bm E \geq 0
\end{equation}
In other words, if $\lambda_{\rm min}(\bm G - \bm E \bm G \bm E) \geq 0$, where $\lambda_{\rm min}(\cdot)$ denotes the minimum eigenvalue of a matrix,  the energy consistency of the impact is ensured. Therefore, the upper-bound limit of CoR can be specified by
\begin{equation}\label{eq:e_star}
e^* := \mbox{arg} \;\; \lambda_{\rm min}(\bm G - \bm E \bm G \bm E),
\end{equation}
regardless of the pre-impact velocity. Notice that \eqref{eq:e_star} gives a conservative upper-bound limit, however  \eqref{eq:e_star}  is independent  of the pre-impact velocity  unlike \eqref{label:quadratic}. From \eqref{eq:mucr_sticking_slipping} and \eqref{eq:e<e_star}, we can say the loci of all CoR and CoF pairs leading to  energetically consistent sticking impacts can be described by the following set:
\begin{equation} \label{eq:consistent_sticking}
\forall e, \mu_s \; \ni \; \mu_s \geq \mu_{\rm cr}(e) \; \wedge \; e \leq e^*
\end{equation}
Suppose $K^-=\frac{1}{2} \dot{\bm q}^{-T}  \bm M \dot{\bm q}^-$ and $K^+=\frac{1}{2} \dot{\bm q}^{+T}  \bm M \dot{\bm q}^+$ are the pre-impact and post-impact values of the kinetic energy of the constrained system. Then, from the expression of the post-impact velocity \eqref{postvel_striction}, one can verify that
\begin{equation}
W_{\rm loss} = K^+ - K^-
\end{equation}

\subsubsection{Slipping mode}
Now consider slip frictional impacts: In this case, the energy loss absorbed in the contact  can be still expressed by \eqref{eq:W_loss}, but the post-impact velocity and impulse are governed by equations \eqref{eq:velocity_postimpact} and \eqref{eq:impulse_slipping}. The average normal and tangential velocities during slipping impact are $\bar v_n = \frac{1}{2}(v_n^+ + v_n^-)=\frac{1}{2}(1-e) \bm a_n \dot{\bm q}^-$ and $\bar{\bm v}_t = \frac{1}{2}\bm A_t (\dot{\bm q}^- + \dot{\bm q}^+ )$, respectively. Thus, the expression of average contact velocity $\bar{\bm v}^T=[\bar{v}_n \; \bar{\bm v}_t^T  ]$  can be written as
\begin{equation} \label{eq:vbar_slipping}
\bar{\bm v} =  \frac{1}{2} \begin{bmatrix} (1-e) \bm a_n \dot{\bm q}^- \\ \bm A_t (\dot{\bm q}^- + \dot{\bm q}^+ ) \end{bmatrix} \quad \Leftarrow \quad  \mu_s < \mu_{\rm cr}
\end{equation}
Upon substitution of \eqref{eq:vbar_slipping} and \eqref{eq:impulse_slipping} in \eqref{eq:loss_slipping}, the expression of power loss during slipping impact can be written as
\begin{align}\notag
W_{\rm loss} &=  -\frac{1}{2} g \Big( (1-e^2)\dot{\bm q}^{-T} \big( \bm a_n^T \bm a_n \big)\dot{\bm q}^{-} - (1+e)  \mu_d \| \bm A_t \bar{\dot{\bm q}} \| \bm a_n \dot{\bm q}^- \Big) \\ \label{eq:loss_slipping}
& =-\frac{1}{2}(1-e^2)g \dot{\bm q}^{-T} \big( \bm a_n^T \bm a_n \big) \dot{\bm q}^{-} + (1+e) g \mu_d \|  \bm A_t \bar{\dot{\bm q}}    \| v_n^-   \leq 0 \quad \Leftarrow \quad  \mu_s < \mu_{\rm cr}
\end{align}
Notice that the first term in the RHS of \eqref{eq:loss_slipping} is always positive semidefinite because $0\leq e \leq 1$. Moreover, the pre-impact normal velocity is always negative, i.e.,
\begin{equation}
\bm a_n \dot{\bm q}^- = v_n^-  < 0
\end{equation}
and hence the second term in the RHS of \eqref{eq:loss_slipping} is always positive definite. Therefore, the condition for negative power losses boils down to the kinematic consistency $g<0$, which has been already stipulated in \eqref{eq:g>0}. Denoting
\begin{equation} \label{eq:mu_star}
\mu^* := \frac{\bm a_n \bm M^{-1} \bm a_n^T}{\| \bm a_n \bm M^{-1} \bm A_t^T \| },
\end{equation}
we can say $g>0$ for all pre-impact velocities  if
\begin{equation}
\quad \mu_s < \mu_{\rm cr}  \quad  \wedge \quad \mu_d < \mu^*
\end{equation}
Therefore, assuming $\mu_s = \mu_d =\mu $, then the loci of energetically consistent CoR and CoF pairs for slipping impact can be described by the following set
\begin{equation} \label{eq:consistent_slipping}
\forall e, \mu \; \ni \; \mu \leq \mbox{min}(\mu_{\rm cr}(e), \mu^*)
\end{equation}
Finally, by virtue of \eqref{eq:consistent_sticking} and \eqref{eq:consistent_slipping} the conditions for energetically consistent of frictional impact during both slipping and sticking modes can be described by
\begin{equation} \label{eq:set_e_mu}
\big\{\mu, e : \; \mu \geq \mu_{\rm cr}(e) \wedge e<e^* \; \cup \; \mu < \mbox{min}(\mu_{\rm cr}(e), \mu^*)   \big\}
\end{equation}
where $\mu_{\rm cr}$, $\mu^*$, and $e^*$ are obtained from \eqref{eq:mucr_e}, \eqref{eq:mu_star}, and \eqref{label:quadratic}  or \eqref{eq:e_star}, respectively. It is important to point out that these variables
are not constant parameters as they change with changing the states and configuration of the MBSs.

\section{Conclusions}
A unifying model slip and stick frictional impact for MBSs has been presented. It has been proven that the model is energetically consistent in both slipping and sticking  modes provided that the values of CoR and CoF are within admissible regions in the variable plan. Unification of the impact model has been  complemented by introducing  a quadratic function of the pre-impact velocity whose sign determines whether the impact occurs in slip state or stick state. This allows  switching to the adequate impact model prior to the impact event. Parametrization of the quadratic function in terms of CoR and CoF variable  led to the Critical CoF, which determines the minimum required CoF to prevent slipping during an impact. Energy consistency  of the impact model in both slipping and sticking friction modes has been analytically investigated and the results revealed that there were upper-bound limits of CoR and CoF and in conjunction with the introduced  Critical CoF variables define the admissible set of CoR and CoF for a consistent impact.

\bibliographystyle{IEEEtran}

\begin{thebibliography}{10}
\bibitem{Muller-1995}
A.~M\"{u}ller, ``Forward dynamics of variable topology mechanisms -- the case
  of constraint activation,'' in \emph{{ECCOMAS} Thematic Conference on
  Multibody Dynamics}, Barcelona, Catalonia, Spain, 2015.

\bibitem{Aghili-2005}
F.~Aghili, ``A unified approach for inverse and direct dynamics of constrained
  multibody systems based on linear projection operator: Applications to
  control and simulation,'' \emph{IEEE Trans. on Robotics}, vol.~21, no.~5, pp.
  834--849, Oct. 2005.

\bibitem{McClamorch-Wang-1988}
N.~H. McClamroch and D.~Wang, ``Feedback stabilization and tracking in
  constrained robots,'' \emph{IEEE Trans. on Automation Control}, vol.~33, pp.
  419--426, 1988.

\bibitem{GarciadeJalon-Bayo-1994}
J.~{Garcia de Jal{\'o}n} and E.~Bayo, \emph{Kinematic and Dynamic Simulation of
  Multibody Systems: The Real-Time Challenge}.\hskip 1em plus 0.5em minus
  0.4em\relax New York: Springer-Verlag, 1994.

\bibitem{Blajer-Schiehlen-Schirm-1994}
W.~Blajer, W.~Schiehlen, and W.~Schirm, ``A projective criterion to the
  coordinate partitioning method for multibody dynamics,'' \emph{Applied
  Mechanics}, vol.~64, pp. 86--98, 1994.

\bibitem{Aghili-2020a}
F.~Aghili, ``Energetically consistent model of slipping and sticking frictional
  impacts in multibody systems,'' \emph{Multibody Syst Dyn}, vol.~48, no.~2,
  pp. 193--209, June 2020.

\bibitem{Preclik-2014}
T.~Preclik, ``Models and algorithms for ultrascale simulations of non-smooth
  granular dynamics,'' Ph.D. dissertation, Technische Fakultat / Department
  Informatik, 2014.

\bibitem{Aghili-Su-2017}
F.~Aghili and C.~Su, ``Impact dynamics in robotic and mechatronic systems,'' in
  \emph{2017 International Conference on Advanced Mechatronic Systems
  (ICAMechS)}, Dec 2017, pp. 163--167.

\bibitem{Gottschlich-Kak-1989}
S.~N. Gottschlich and A.~C. Kak, ``A dynamic approach to high-precision parts
  mating,'' \emph{IEEE Transactions on Systems, Man, and Cybernetics}, vol.~19,
  no.~4, pp. 797--810, Jul 1989.

\bibitem{Walker-1994}
I.~D. Walker, ``Impact configurations and measures for kinematically redundant
  and multiple armed robot systems,'' \emph{IEEE Transactions on Robotics and
  Automation}, vol.~10, no.~5, pp. 670--683, Oct 1994.

\bibitem{Aghili-2010h}
F.~Aghili, ``Control of redundant mechanical systems under equality and
  inequality constraints on both input and constraint forces,'' \emph{{ASME}
  Journal of Computational and Nonlinear Dynamics}, vol.~6, no.~3, July 2011.

\bibitem{Dupree-Liang-2008}
K.~Dupree, C.~H. Liang, G.~Hu, and W.~E. Dixon, ``Adaptive lyapunov-based
  control of a robot and mass-spring system undergoing an impact collision,''
  \emph{IEEE Transactions on Systems, Man, and Cybernetics, Part B
  (Cybernetics)}, vol.~38, no.~4, pp. 1050--1061, Aug 2008.

\bibitem{Su-Stepanenko-1995}
C.-Y. Su and Y.~Stepanenko, ``Adaptive sliding mode coordinated control of
  multiple robot arms attached to a constrained object,'' \emph{IEEE
  Transactions on Systems, Man, and Cybernetics}, vol.~25, no.~5, pp. 871--878,
  May 1995.

\bibitem{Aghili-Buehler-Hollerbach-1997a}
F.~Aghili, M.~Buehler, and J.~M. Hollerbach, ``Dynamics and control of
  direct-drive robots with positive joint torque feedback,'' in \emph{{IEEE}
  Int. Conf. Robotics and Automation}, vol.~11, 1997, pp. 1156--1161.

\bibitem{Zhang-Xi-2009}
F.~Zhang, Y.~Xi, Z.~Lin, and W.~Chen, ``Constrained motion model of mobile
  robots and its applications,'' \emph{IEEE Transactions on Systems, Man, and
  Cybernetics, Part B (Cybernetics)}, vol.~39, no.~3, pp. 773--787, June 2009.

\bibitem{Aghili-2015b}
F.~Aghili, ``Non-minimal order model of mechanical systems with redundant
  constraints for simulations and controls,'' \emph{IEEE Transactions on
  Automatic Control}, vol.~61, no.~5, pp. 1350--1355, May 2016.

\bibitem{Marhefka-Orin-1999}
D.~W. Marhefka and D.~E. Orin, ``A compliant contact model with nonlinear
  damping for simulation of robotic systems,'' \emph{IEEE Transactions on
  Systems, Man, and Cybernetics - Part A: Systems and Humans}, vol.~29, no.~6,
  pp. 566--572, Nov 1999.

\bibitem{Mu-Wu-2006}
X.~Mu and Q.~Wu, ``On impact dynamics and contact events for biped robots via
  impact effects,'' \emph{IEEE Transactions on Systems, Man, and Cybernetics,
  Part B (Cybernetics)}, vol.~36, no.~6, pp. 1364--1372, Dec 2006.

\bibitem{Konno-Myojin-2011}
A.~Konno, T.~Myojin, T.~Matsumoto, T.~Tsujita, and M.~Uchiyama, ``An impact
  dynamics model and sequential optimization to generate impact motions for a
  humanoid robot,'' \emph{The International Journal of Robotics Research,},
  vol.~30, no.~13, pp. 1596--1608, 2011.

\bibitem{Yoshida-Takeuchi-2014}
Y.~Yoshida, K.~Takeuchi, Y.~Miyamoto, D.~Sato, and D.~Nenchev, ``Postural
  balance strategies in response to disturbances in the frontal plane and their
  implementation with a humanoid robot,'' \emph{IEEE Transactions on Systems,
  Man, and Cybernetics: Systems}, vol.~44, no.~6, pp. 692--704, June 2014.

\bibitem{Nenchev-Yoshida-1998}
D.~N. {Nenchev} and K.~{Yoshida}, ``Impact analysis and post-impact motion
  control issues of a free-floating space robot subject to a force impulse,''
  \emph{IEEE Transactions on Robotics and Automation}, vol.~15, no.~3, pp.
  548--557, June 1999.

\bibitem{Keller-1986}
J.~B. Keller, ``Impact with friction,'' \emph{Journal of Applied Mechanics},
  vol.~53, no.~1, pp. 1--4, 1986.

\bibitem{Glocker-Pfeiffer-1995}
C.~Glocker and F.~Pfeiffer, ``Multiple impacts with friction in rigid multibody
  systems,'' \emph{Nonlinear Dynamics}, vol.~7, pp. 471--497, Jun. 1995.

\bibitem{Stronge-James-Ravani-2001}
W.~J. Stronge, R.~James, and B.~Ravani, ``Oblique impact with friction and
  tangential compliance,'' \emph{Philosophical Transactions: Mathematical,
  Physical and Engineering Sciences}, vol. 359, no. 1789, pp. 2447--2465, 2001.

\bibitem{Johansson-2001}
L.~Johansson, ``A newton method for rigid body frictional impact with multiple
  simultaneous impact points,'' \emph{Computer Methods in Applied Mechanics and
  Engineering}, no. 191, pp. 239--254, 2001.

\bibitem{Pfeiffer-2009}
F.~Pfeiffer, \emph{Mechanical System Dynamics}.\hskip 1em plus 0.5em minus
  0.4em\relax Springer, 2009.

\bibitem{Glocker-2013}
C.~Glocker, ``Energetic consistency conditions for standard impacts,''
  \emph{Multibody Syst. Dyn.}, vol.~29, no.~1, pp. 77--117, 2013.

\bibitem{Moreau-1988}
J.~J. Moreau, \emph{Topics in Nonsmooth Mechanics}.\hskip 1em plus 0.5em minus
  0.4em\relax Birkhauser, 1981, ch. Bounded Variation in Time, pp. 1--74.

\bibitem{Brogliato-1999}
B.~Brogliato, \emph{Nonsmooth mechanics, 2nd edn.}\hskip 1em plus 0.5em minus
  0.4em\relax London, UK: Springer, 1999.

\bibitem{Acary-Brogliato-2008}
V.~Acary and B.~Brogliato, \emph{Numerical Methods for Nonsmooth Dynamical
  Systems: Applications in Mechanics and Electronics}.\hskip 1em plus 0.5em
  minus 0.4em\relax Springer, 2008.

\bibitem{Leine-2008}
R.~I. Leine and N.~van~de Wouw, \emph{Stability and Convergence of Mechanical
  Systems with Unilateral Constraints}.\hskip 1em plus 0.5em minus 0.4em\relax
  Springer, 2008.

\bibitem{Alart-Curnier-1991}
P.~Alart and A.~Curnier, ``A mixed formulation for frictional contact problems
  prone to newton like solution methods,'' \emph{Comput Methods Appl Mech Eng},
  vol.~92, pp. 353--375, 1991.

\bibitem{Schindler-Nguyen-2011}
T.~{Schindler}, B.~{Nguyen}, and J.~{Trinkle}, ``Understanding the difference
  between prox and complementarity formulations for simulation of systems with
  contact,'' in \emph{2011 IEEE/RSJ International Conference on Intelligent
  Robots and Systems}, Sep. 2011, pp. 1433--1438.

\bibitem{Stewart-2011}
D.~E. Stewart, \emph{Dynamics with Inequalities: Impacts and Hard
  Constraints}.\hskip 1em plus 0.5em minus 0.4em\relax New York, United States:
  Society for Industrial \& Applied Mathematics, 2011.

\bibitem{Zheng-Hemami-1985}
Y.~F. Zheng and H.~Hemami, ``Mathematical modeling of a robot collision with
  its environment,'' \emph{Journal of Robotic Systems}, vol.~1, pp. 289--307,
  1985.

\bibitem{Volpe-Khosla-1993}
R.~Volpe and P.~Khosla, ``A theoretical and experimental investigation of
  impact control for manipulators,'' \emph{The International Journal of
  Robotics Research}, vol.~21, pp. 351--365, 1993.

\bibitem{Stronge-2000}
W.~J. Stronge, \emph{Impact Mechanics}.\hskip 1em plus 0.5em minus 0.4em\relax
  Cambridge University Press, 2000.

\bibitem{Khulief-2013}
Y.~A. Khulief, ``Modeling of impact in multibody systems: An overview,''
  \emph{ASME Journal of Computational and Nonlinear Dynamics}, vol.~8, April
  2013.

\bibitem{Hunt-Crossley-1975}
K.~H. Hunt and F.~R.~E. Crossley, ``Coefficient of restitution interpreted as
  damping in vibroimpact,'' vol.~42, pp. 440--445, Jun. 1975, series E.

\bibitem{Goldsmith-1960}
W.~Goldsmith, \emph{Impact: The Theory and Physical Behavior of Colliding
  Solids}.\hskip 1em plus 0.5em minus 0.4em\relax London, U.K.: Edward Arnol,
  1960.

\bibitem{Lankarani-Nikravesh-1990}
H.~M. Lankarani and P.~E. Nikravesh, ``A contact force model with hysteresis
  damping for impact analysis of mutlibody systems,'' \emph{ASME Journal of
  Mechanical Design}, vol. 112, pp. 369--376, Sep. 1990.

\bibitem{Aghili-2019a}
F.~Aghili, ``Modelling and analysis of multiple impacts in multibody systems
  under unilateral and bilateral constrains based on linear projection
  operators,'' \emph{Multibody Syst. Dyn.}, vol.~46, no.~1, pp. 41--62, 2019.

\bibitem{Kane-Levinson-1985}
T.~Kane and D.~Levinson, \emph{Dynamics: theory and applications}.\hskip 1em
  plus 0.5em minus 0.4em\relax New York: McGraw-Hill, 1985.

\bibitem{Wang-Mason-1992}
Y.~W. Y and M.~Mason, ``Two-dimensional rigid-body collisions with friction,''
  \emph{Journal of Applied Mechanics}, vol.~59, pp. 635--642, 1992.

\bibitem{Djerassi-2009}
S.~Djerassi, ``Collision with friction, part a,'' \emph{Multibody Syst. Dyn.},
  vol.~21, pp. 37--54, 2009.

\bibitem{Jia-2013}
Y.-B. Jia, ``Three-dimensional impact: energy-based modeling of tangential
  compliance,'' \emph{The International Journal of Robotics Research}, vol.~32,
  no.~1, pp. 56--83, 2013.

\bibitem{Chatterjee-Ruina-1988}
A.~Chatterjee and A.~Ruina, ``A new algebraic rigid-body collision law based on
  impulse space considerations.'' \emph{Journal of Applied Mechanics}, vol.~65,
  pp. 939--951, 1988.

\bibitem{Stewart-Trinkle-1996}
D.~E. Stewart and J.~C. Trinkle, ``An implicit time-stepping scheme for rigid
  body dynamics with inelastic collisions and coulomb friction.''
  \emph{International Journal for Numerical Methods in Engineering}, vol.~39,
  pp. 2673--2691, 1996.

\bibitem{Nguyen-Brogliato-2013}
N.~S. Nguyen and B.~Brogliato, \emph{Multiple Impacts in Dissipative Granular
  Chains}.\hskip 1em plus 0.5em minus 0.4em\relax Springer, 2013.

\bibitem{Flickinger-Bowling-2009}
D.~M. Flickinger and A.~Bowling, ``Simultaneous oblique impacts and contacts in
  multibody systems with friction,'' in \emph{2009 IEEE/ASME International
  Conference on Advanced Intelligent Mechatronics}, July 2009, pp. 1613--1618.

\bibitem{Stronge-1991}
W.~Stronge, ``Unraveling paradoxical theories for rigid body collisions,''
  \emph{Journal of Applied Mechanics}, vol.~58, pp. 1049--1055, 1991.

\bibitem{Stoianovici-Hurmuzlu-1996}
D.~Stoianovici and Y.~Hurmuzlu, ``A critical study of the applicability of
  rigid-body collision theory,'' \emph{ASME Journal of Appl. Mech.}, vol.~63,
  no.~2, pp. 307--316, 1996.

\end{thebibliography}

\end{document}